\documentclass[sigconf]{acmart}
\usepackage{booktabs}

\setcopyright{none}
\renewcommand\footnotetextcopyrightpermission[1]{}

\begin{document}

\title{Standard-Anchored Delegation in Teachers’ Use of Generative AI}
\author{Muhammad Saleh Farooq}
\affiliation{\institution{Lahore University of Management Sciences}\city{Lahore}\country{Pakistan}}
\author{Muhammad Ahmad Khan}
\affiliation{\institution{Lahore University of Management Sciences}\city{Lahore}\country{Pakistan}}
\author{Muhammad Jon Raza}
\affiliation{\institution{Lahore University of Management Sciences}\city{Lahore}\country{Pakistan}}
\renewcommand{\shortauthors}{Farooq et al.}

\begin{abstract}
Teachers who use generative AI must decide which parts of their work to hand over, while they remain accountable for what reaches students. We studied these decisions in Punjab, Pakistan, through a qualitatively driven mixed-methods study. We interviewed 13 teachers from government, private, NGO and vocational institutions, four administrators and one student, and coded each teacher's position on twelve educational tasks. Nine teachers accepted AI drafts or substantial assistance with lesson planning, but none accepted AI-drafted learning objectives, and eleven did not use AI to encourage or emotionally support students. Teachers who would accept AI marking required that the model be given a rubric or marking criteria. Building on assessment research and HCI work on verifiability, we call this pattern \emph{standard-anchored delegation}: teachers handed a component of a task to AI when a standard outside the model, such as a textbook or rubric, defined a correct output and the teacher could apply that standard to the result. Having a standard was not always enough. Several government-school teachers did not use AI to mark quizzes even though the quizzes had answer keys, because they expected students to look for evidence of the teacher's effort. Checking AI output required subject knowledge and time, and private schools gave teachers more support for this than government schools did. Participants also read disclosure of AI use as a statement about whether the teacher had reviewed the work. We propose that teacher-facing AI tools ask for the relevant standard before generating and show each output beside it, and that schools account for the time that checking takes.
\end{abstract}



\maketitle
\pagestyle{plain}

\section{Introduction}

Teachers who use generative AI remain accountable for what it produces. A question paper that strays from the prescribed textbook, or a comment that misreads a struggling student, is the teacher's error once it reaches the class. Every act of delegation therefore carries a second task: checking the output before anyone uses it. This paper examines how teachers decide what to delegate when they must check what comes back.

HCI research has examined delegation to AI in two main forms. Studies of task delegability ask which tasks people want AI to perform, and find that motivation, difficulty, risk and trust shape those preferences~\cite{lubars2019ask}. Studies of reliance ask when people accept or override AI recommendations in decisions with a correct answer, and how explanations or cognitive forcing change that behaviour~\cite{bucinca2021trust,vasconcelos2023explanations,lee2004trust}. Teaching fits neither form well. Teachers rarely hand over a whole task, and a lesson plan or feedback comment has no ground-truth label. Its quality depends on a curriculum and on what the teacher knows about particular students, and the teacher establishes it after generation. Education research describes how teachers make such judgements. Judging a piece of work requires a standard and the expertise to compare the work with it~\cite{sadler1989formative}, and curriculum design aligns activities and assessments to intended learning outcomes~\cite{biggs1996enhancing}. We bring these accounts to teachers' decisions about AI.

Research on teachers and AI documents teachers' demand for control over AI recommendations~\cite{holstein2019codesigning,lawrence2024shared}, their adaptation of generated resources to learners~\cite{jeon2023llm}, and curated deployments in which reviewers check AI-generated lesson plans before teachers customise them~\cite{dennison2026shiksha}. These studies largely examine purpose-built tools. Many teachers in the Global South meet generative AI through general-purpose chatbots on their own phones, inside institutions with uneven resources and few shared rules~\cite{raza2025pakistani}. How these teachers allocate responsibility between themselves and AI, and what makes the necessary checking possible, has received little study.

We examine these questions in Punjab, Pakistan, where government schools, low-fee and elite private schools, NGO schools and vocational institutes operate under different curricula and administrative demands~\cite{andrabi2008dime,aser2024,nasreen2019time}. Our qualitatively driven mixed-methods study combines interviews with 18 teachers, administrators and a student with ordinal codes for teachers' positions on twelve educational tasks. We ask:

\begin{description}
\item[RQ1] How do teachers divide educational work between themselves and generative AI, and what governs the division?
\item[RQ2] What conditions shape the checking that delegation requires, including how teachers make that checking visible to students and parents?
\end{description}

Teachers delegated the production of material and kept the standards that material had to meet. Nine of 13 teachers accepted drafts or substantial assistance with lesson planning; eight of those nine limited AI to formatting or suggesting the plan's learning objectives. Teachers who explained when they would accept AI marking tied it to a rubric given to the model, and eleven kept encouragement for themselves. Grading also split by sector, and government-school teachers who refused it explained the refusal through the effort students expected to see. Checking depended on curricular artefacts and subject expertise, and institutions left very different amounts of time for it. Teachers and the student participant treated disclosure of AI assistance as information about whether the teacher had reviewed the work.

We make two contributions. First, we describe \emph{standard-anchored delegation}, grounded in task-level evidence from government, private, NGO and vocational institutions: teachers delegate a component of work when a standard outside the model defines a correct output and the teacher can state that standard and apply it to the result. The concept joins HCI research on verifiability with assessment research on standards and alignment. It explains why the same teacher delegated a lesson's activities and kept its objectives, it places the capacity to verify with the teacher and the institution as well as with the output, and it identifies where verifiability stops explaining teachers' choices. Second, we derive implications for teacher-facing AI and school policy that treat verification as work to support, budget and report.

\section{Related Work}

Our study draws on research on delegation and reliance in human--AI interaction, on assessment standards and teachers' work with AI, on AI-mediated communication, and on AI in under-resourced education systems.

\subsection{Delegation and Reliance }

HCI and machine-learning research has approached delegation to AI as a question of which tasks people will hand over. Lubars and Tan~\cite{lubars2019ask} surveyed preferences across a hundred tasks and found that motivation, difficulty, risk and trust shape delegability, with people often preferring AI assistance to full automation. Work on reliance studies what happens once a person receives AI output. Lee and See~\cite{lee2004trust} define appropriate reliance as trust calibrated to a system's capability, and experimental studies of AI-assisted decisions operationalise it as accepting correct recommendations and rejecting incorrect ones~\cite{bucinca2021trust,vasconcelos2023explanations}. Bu\c{c}inca et al.\ show that cognitive forcing functions reduce overreliance at a cost to user satisfaction~\cite{bucinca2021trust}, and Vasconcelos et al.\ show through a cost-benefit framework that people decide strategically whether to verify AI output, and that the difficulty of the task and of the explanation both shape overreliance~\cite{vasconcelos2023explanations}. Fok and Weld argue that explanations help only when they let people verify AI output~\cite{fok2024verifiability}. Shneiderman's framework for human-centred AI treats automation and human control as separate dimensions, so that a system can automate heavily while people retain control~\cite{shneiderman2020hcai}.

Generative AI changes what verification involves. Its outputs are artefacts that users must evaluate, and Tankelevitch et al.\ describe the metacognitive demands this places on users, who must know what a good output looks like and monitor the result~\cite{tankelevitch2024metacognitive}. Simkute et al.\ extend Bainbridge's ironies of automation~\cite{bainbridge1983ironies} to generative AI, showing how a shift from production to evaluation can erode productivity gains~\cite{simkute2024ironies}. A survey of knowledge workers found that generative AI moves effort from producing work to verifying and integrating it~\cite{lee2025critical}, and Risko and Gilbert's account of cognitive offloading explains why such moves reduce some demands while creating others~\cite{risko2016offloading}. Tankelevitch et al.\ already treat task decomposition and output evaluation as central to generative AI use, and Fok and Weld place verification at the centre of reliance. The experimental studies of reliance assume a correct answer against which acceptance can be scored. We examine work without such an answer, in which teachers judge each component of a task against a standard such as a textbook or their knowledge of a student, and we ask where that standard comes from and who holds it.

\subsection{Teachers, Standards and AI}

Assessment research describes how teachers judge work. Sadler argues that judging the quality of a piece of work requires a concept of the standard it should meet, the ability to compare the work with that standard and a way to act on the difference~\cite{sadler1989formative}. Tai et al.\ call the capability to make such decisions evaluative judgement~\cite{tai2018evaluative}. Curriculum design places standards in sequence. Constructive alignment derives teaching activities and assessments from intended learning outcomes~\cite{biggs1996enhancing}, and backward design asks teachers to set the desired results before planning the activities that lead to them~\cite{wiggins2005understanding}. Shulman's account of the knowledge base for teaching separates knowledge of subject matter from knowledge of learners and their characteristics~\cite{shulman1987knowledge}. These accounts describe the standards teachers apply to their own and their students' work; we examine how teachers apply them to AI output.

Research on teacher--AI collaboration has documented teachers' demand for control. Holstein et al.\ co-designed a classroom orchestration tool with teachers and framed teacher--AI complementarity around the teacher's authority to act on AI information~\cite{holstein2019codesigning}. Lawrence et al.\ followed teachers through a multiyear design process for an AI tool that pairs students and found that teachers wanted to configure and override its recommendations~\cite{lawrence2024shared}. With generative AI, Jeon and Lee found that elementary English teachers used ChatGPT to produce material at different levels while retaining pedagogical decisions~\cite{jeon2023llm}. Woodrow et al.\ built teaching assistants' selections and edits of AI feedback into model training~\cite{woodrow2025feedback}, and a large trial in England measured the time teachers saved when using ChatGPT for lesson preparation~\cite{roy2024chatgpt}.

Closest to our setting, Dennison et al.\ studied Shiksha Copilot in government schools in Karnataka, where curators reviewed AI-generated lesson plans before teachers customised them~\cite{dennison2026shiksha}. Their study shows the value of a curation layer between model and classroom. We study teachers who adopt general-purpose chatbots without such a layer, and we examine the full range of tasks from planning to emotional support, which lets us compare where teachers drew boundaries within the same body of work.

\subsection{AI-Mediated Communication and Disclosure}

Hancock et al.\ define AI-mediated communication as interpersonal communication in which an intelligent agent modifies, augments or generates messages on a communicator's behalf~\cite{hancock2020aimc}. Experimental work in this area finds that recipients judge text they believe AI wrote as less trustworthy~\cite{jakesch2019aimc} and that disclosed AI involvement lowers trust in a sender, whose use of AI recipients read as low effort and insincerity~\cite{liu2022console}. The penalty matches a broader finding that people judge work by the effort they believe went into it~\cite{kruger2004effort}, and that showing the work behind a service raises its perceived value~\cite{buell2011labor}. Users of AI writing tools consider them acceptable for formal messages and inauthentic for close ones~\cite{fu2024text}. Draxler et al.\ find that writers feel little ownership of AI-generated text yet still declare themselves its authors~\cite{draxler2024ghostwriter}. Feedback on students' work and messages to parents are AI-mediated communication with an institutional dimension: the teacher speaks for the school and is accountable for the assessment the message conveys. We examine how teachers, administrators and a student reason about disclosure in this setting, and what they take disclosure to reveal.

\subsection{AI in Under-Resourced Education Systems}

HCI research in the Global South has shown that frontline workers' encounters with AI depend on the institutions around them. Okolo et al.\ found that community health workers in rural India anticipated misdiagnosis, surveillance and loss of trust when imagining an AI diagnostic application~\cite{okolo2021cannot}. Kapania et al.\ found that Indian users granted AI authority without evidence of its capability, which made them more tolerant of its errors~\cite{kapania2022because}, and Ramjee et al.\ found that the nurses answering questions escalated by an LLM chatbot for health workers worried about their own workload and accountability~\cite{ramjee2025ashabot}. Sambasivan et al.\ argue that fairness frameworks developed in the West omit the institutional conditions of South Asia~\cite{sambasivan2021reimagining}. Pakistan's schooling spans government schools, a large low-fee private sector and elite private schools~\cite{andrabi2008dime,aser2024}, and government teachers carry non-teaching duties that compete with instruction~\cite{nasreen2019time}. A survey of 125 Pakistani educators found that respondents used AI more for generating content than for grading or feedback and raised concerns about unequal access~\cite{raza2025pakistani}, and a training intervention in a Karachi government school supported teachers through bilingual handouts and a WhatsApp group~\cite{hafeez2025assistant}. We extend this work by examining how teachers across these sectors allocate responsibility to AI and how their institutions shape the checking that allocation requires.

\section{Method}
\label{sec:method}

We conducted a qualitatively driven mixed-methods study~\cite{morse2009mixed,creswell2018designing}. The core qualitative component analysed semi-structured interviews with 18 participants. The supplementary quantitative component converted each teacher's interview account into ordinal delegation codes for twelve educational tasks and summarised their distribution. We integrated the two components at the level of tasks and participants, using the codes to locate agreement and divergence and the interviews to explain them.

\subsection{Setting and Participants}
\label{sec:setting}

Punjab is Pakistan's most populous province, and its schooling spans government schools, low-fee and elite private schools, NGO schools and vocational institutes~\cite{andrabi2008dime,aser2024}. We recruited participants in Lahore, Multan and surrounding urban and rural areas through purposive sampling and institutional referrals, seeking variation in sector, role and teaching conditions. At the time of the study, the Punjab School Education Department had issued no official guidelines on teachers' use of generative AI. Pakistan's national AI policy addresses capacity building~\cite{moitt2025policy}, and the Cambridge and IB examination boards regulate students' use of AI~\cite{cambridge_genai,ib2023statement}. Table~\ref{tab:participants} describes the 18 participants: 13 teachers, four administrators and one student. Six teachers worked in government schools and five in private schools; P1 taught and coordinated teachers in an NGO school, and P16 taught physics in a vocational institute. Teachers' experience ranged from three to 24 years, and their class sizes from about 6 to 70 pupils. The administrators led colleges and a vocational institute, and P2 was a student in a private school.

\begin{table*}[t]
\caption{Participants. Experience in years; administrators' experience refers to administration where reported. Gov.\ = government; Priv.\ = private; Voc.\ = vocational; -- = not reported.}
\label{tab:participants}
\small
\begin{tabular}{@{}lllllll@{}}
\toprule
ID & Role & Sector & Area & Subject and grades & Exp. & Class size\\
\midrule
P1 & Teacher, coordinator & NGO & Urban & English, sciences; 5--9 & 3 & $\sim$20\\
P2 & Student & Priv. & Urban & -- & -- & --\\
P3 & Teacher & Gov. & Urban & Mathematics; 9, 12 & 24 & 50--70\\
P4 & Teacher (several schools) & Priv. & Urban & Sciences; O/A-level, IGCSE, IB & 17 & 25--30\\
P5 & Teacher, examinations lead & Priv. & Urban & Computer science; 11--12 (IB) & 10 & 6--12\\
P6 & Teacher & Priv. & Urban & Biology; 9--12 & 10 & 22--28\\
P7 & Teacher & Priv. & Urban & Mathematics; 9--12 & 14.5 & $\sim$25\\
P8 & Principal & College & Rural & -- & 13 & --\\
P9 & Teacher & Priv. & Urban & Physics; O-level & 10 & $\sim$20\\
P10 & Teacher & Gov. & Rural & Several subjects & 8.5 & $\sim$30\\
P11 & Teacher & Gov. & Rural & Several subjects; 1--8 & 12 & $\sim$30\\
P12 & Teacher & Gov. & Rural & Social studies & 14 & $\sim$30\\
P13 & Teacher & Gov. & Rural & Several subjects & 7 & $\sim$30\\
P14 & Principal & College & Urban & -- & $>$20 & --\\
P15 & Principal & Voc.\ institute & Rural & -- & $>$20 & --\\
P16 & Senior teacher & Voc.\ institute & Rural & Physics & $>$20 & --\\
P17 & Teacher, master trainer & Gov. & Urban & Computer science & 19 & --\\
P18 & Vice-principal & College & Urban & -- & $>$10 & --\\
\bottomrule
\end{tabular}
\end{table*}

\subsection{Interviews}

Three authors conducted the interviews in Urdu, English or a mix of both between March and October 2026. The first seven interviews (P1--P7, March--April 2026) served as the pilot. We administered the same role-specific protocols throughout the study, so the pilot interviews form part of the 18-participant sample. The protocol planned 30--40 minutes for teacher interviews and 20--30 minutes for administrator and student interviews, and asked each participant's permission to audio-record. The teacher protocol covered participants' roles and workload, their current use of AI, and eleven task prompts spanning preparation, assessment, feedback and communication with parents. For each task, we asked whether the teacher would use AI, what form the assistance would take, what checking or editing the output would need and whom an error would affect. Follow-up questions compared tasks and probed factual accuracy, emotional suitability, time savings and student vulnerability.

Two vignettes elicited views on disclosure. In the first, a teacher uses AI to draft personalised feedback on a weak essay and edits it before sharing; in the second, a teacher uses AI to help mark or summarise performance on a short quiz, then checks it. We asked whether students should be told and how students or parents would judge the teacher's effort and credibility. The administrator protocol addressed institutional positions on AI, acceptable and risky uses, and disclosure; the student protocol addressed meaningful feedback, perceived teacher effort and trust. The authors translated Urdu segments into English. Text in quotation marks reproduces participants' words, in our translation where they spoke Urdu, lightly edited for readability; other attributions are close paraphrases.
Interviews took place in person or over Zoom. We transcribed the recordings, which mixed Urdu and English, with automated multilingual speech-recognition models.

\subsection{Analysis}

\paragraph{Qualitative component.} We analysed the interviews through reflexive thematic analysis~\cite{braun2006thematic,braun2019reflecting}. For each participant we prepared a case record of AI uses, task positions, reasons and institutional conditions, drawn from thematic coding sheets or, for P15 and P16, directly from transcripts. Each coding sheet organised coded passages under broad themes and more specific codes; the seven pilot interviews alone yielded more than 340 coded passages. We assembled the case records into a participant-by-theme matrix and rated each theme's salience in each case as strong, moderate or weak. Appendix Table~\ref{tab:themes} reports the strong ratings by sector; the ratings summarise the coding record and do not establish that a theme was absent elsewhere. Comparing accounts of the same task across participants, we developed themes around the components of work teachers retained, the standards they checked outputs against and the institutional arrangements surrounding that checking. We treated cases that departed from an emerging pattern as tests of it; P13's acceptance of AI-assisted grading and the government-school refusals discussed in Section~\ref{sec:f-standard} shaped how we bounded the central concept.

\paragraph{Quantitative component.} We quantitised~\cite{sandelowski2009quantitizing} each teacher's position on twelve task categories into four ordinal codes: no AI (A), ideas or checking only (B), a draft that the teacher edits or reviews (C), and substantial or near-final assistance (D). The categories combine the protocol's task prompts and the quiz-marking vignette with uses participants described when discussing their current practice, such as lesson planning and presentations. We coded a category as unclear when the interview did not establish a position. Unclear codes concentrate among government-school teachers whose interviews covered fewer categories (P10, P11), and we report them in every distribution. We retained one mixed code where P13 described different positions for different parent messages. The 156 teacher--task cells yield the distributions in Table~\ref{tab:joint}, the sector breakdown in Table~\ref{tab:sector} and the individual codes in Appendix~\ref{app:codes}.
One author (M.J.R.) coded all interviews, prepared the case records and assigned the task codes. Reflexive thematic analysis treats the analyst's interpretation as part of the method, so we did not calculate inter-coder agreement~\cite{braun2019reflecting}.

\paragraph{Integration.} We built a joint display (Table~\ref{tab:joint}) that places each task's code distribution beside the standard against which participants described checking its output. The display worked in both directions: divergent distributions sent us to the interview passages that explained them, and recurring explanations sent us to compare tasks that shared a standard. Within-participant comparisons, such as the same teacher's codes for lesson planning and learning objectives, carried most weight in this integration.

\subsection{Positionality and Ethics}

Our team combines HCI and education research at a university in Lahore, and the interviewers share participants' languages. Our familiarity with Pakistan's school sectors shaped which institutional conditions we probed and how we read accounts of workload and oversight. We report participants by identifier and omit the names of institutions.The study was approved by the Institutional Review Board of the Lahore University of Management Sciences (LUMS; IRB registration IORG0007532). Before each interview we went through a consent form that explained the study's purpose and procedures, that participation was voluntary, that participants could withdraw at any time without consequence, and how we would store and anonymise the data. Participants then gave verbal informed consent in English or Urdu, which is recorded in the transcripts, and consented separately to audio recording. Participants received no compensation. We assigned each participant an identifier when we collected their data, removed identifying details during transcription and deleted the audio recordings once transcription was complete. We report participants by identifier and omit the names of institutions.

\section{Findings}
\label{sec:findings}

Sections~\ref{sec:f-production} and~\ref{sec:f-standard} address RQ1, and Sections~\ref{sec:f-expertise} to~\ref{sec:f-disclosure} address RQ2. Counts refer to the 13 teachers unless we state otherwise. The codes and accounts describe what participants reported doing, or said they would do when responding to the task prompts and vignettes; we did not observe their practice. Appendix Tables~\ref{tab:sector} and~\ref{tab:themes} break the task codes and theme ratings down by sector. Table~\ref{tab:joint} places each task's codes beside the standard against which participants described checking its output: a codified standard such as a textbook or rubric, the teacher's knowledge of particular learners, or the learning objectives the teacher set.

\begin{table*}[t]
\caption{Joint display of teacher task codes ($n=13$) and the standards against which participants described checking each output. A = no AI; B = ideas or checking only; C = draft edited or reviewed; D = substantial or near-final assistance; A/D = mixed code (P13, who described different positions for different parent messages); ? = unclear. Codified = textbook, syllabus, rubric or answer key; learners = the teacher's knowledge of particular students or classes. Participant IDs mark accounts that describe the check. We order categories by the number of C and D codes; data analysis and policy appear in Appendix~\ref{app:codes}.}
\label{tab:joint}
\small
\begin{tabular}{@{}lrrrrrrlp{5.2cm}@{}}
\toprule
Task category & A & B & C & D & A/D & ? & Standard & Standard used to check the output (participants)\\
\midrule
Lesson planning & 0 & 4 & 4 & 5 & 0 & 0 & Codified & Syllabus scope (P9)\\
Quiz and test generation & 0 & 4 & 6 & 3 & 0 & 0 & Codified & Textbook chapter (P17)\\
Presentations and content & 0 & 6 & 5 & 2 & 0 & 0 & Codified & Syllabus depth, read through subject knowledge (P9)\\
Worksheets and differentiation & 0 & 2 & 6 & 1 & 0 & 4 & Codified, learners & Pupils' attainment and language (P13)\\
Quiz grading & 3 & 3 & 5 & 1 & 0 & 1 & Codified & Rubric or answer key given to the model (P5, P16)\\
Student feedback & 2 & 1 & 6 & 0 & 0 & 4 & Codified, learners & Rubric and the teacher's reading of the work (P2, P5)\\
Report-card comments & 1 & 2 & 4 & 0 & 0 & 6 & Learners & The teacher's knowledge of the student's progress\\
Parent communication & 2 & 2 & 3 & 0 & 1 & 5 & Learners & Relationship with the family (P7)\\
Encouragement and emotional support & 11 & 0 & 2 & 0 & 0 & 0 & Learners & The student's circumstances (P1, P4, P7)\\
Learning objectives & 1 & 10 & 0 & 0 & 0 & 2 & Teacher-set & The teacher's goals for the class (P1)\\
\bottomrule
\end{tabular}
\end{table*}

\subsection{Teachers Delegated Production and Kept the Standards}
\label{sec:f-production}

Preparation was the most delegated work in the sample. Nine teachers accepted drafts or substantial assistance with lesson planning, and nine accepted the same for quizzes and tests. P4 dictated lesson plans to ChatGPT through voice messages while commuting between schools. P5 and P6 used institutional lesson-planning tools that organised plans around Bloom's taxonomy, and P6 reported that one such platform produced more than a thousand plans a day. In government institutions, P3 used AI to generate mathematics questions and P17 to draft question papers.

Learning objectives drew a different response from the same teachers. Our prompt asked about using AI to format lesson objectives. Ten teachers limited AI to formatting, suggesting ideas or checking wording, P1 excluded it, and none accepted an AI draft. The comparison holds within individuals: of the nine teachers who delegated lesson planning, we coded eight A or B on objectives and one as unclear. P1 described the division inside a single plan. AI could propose activities and draft the sequence, while the teacher decided what pupils should learn and how the lesson would run. P3 expected teachers to understand material during preparation so that they could explain it themselves in class. P12 framed the risk as effort, warning that ready-made answers could make ``your own effort becomes zero.''

A lesson plan thus contained two kinds of component. The activities, explanations and questions were material that teachers could inspect against a syllabus. The objectives were the standard against which the teacher inspected them, and teachers treated the choice of that standard as theirs. This is the order curriculum design prescribes, with outcomes set first and activities planned to serve them~\cite{biggs1996enhancing,wiggins2005understanding}.

\subsection{Teachers Delegated Where a Standard Existed, Unless Students Expected to See Their Effort}
\label{sec:f-standard}

Grading divided teachers more than any preparatory task. Six accepted drafting or substantial assistance with quiz checking, three restricted AI to ideas or checking, and three excluded it. The division followed sector: four of five private-school teachers held C or D codes, against one of six government-school teachers. The accepting teachers who explained their condition tied it to a standard handed to the model. P5 accepted AI marking for multiple-choice questions and, for other work, only when the full rubric accompanied the request, and held to that threshold under pressure:

\begin{quote}
Even if I'm running out of time and having work pressure, if AI isn't able to serve the purpose, what's the point of using it? (P5, private school)
\end{quote}

\noindent P16, in the vocational institute, gave the model grading criteria and adopted its suggestions according to their feasibility, and P15, an administrator, allowed AI grading only if the teacher made sure it was marking correctly.

A quiz's answer key gave the government-school teachers who refused AI marking a codified standard of the same kind. Their reasons concerned what students would see. P3 classed checking as the teacher's own work and expected students to read AI-assisted checking as a shortcut. P12 placed quiz checking among the tasks where the teacher's effort had to remain, and P17 kept grading for himself. P11, who limited AI to checking, expected students to conclude that a teacher who disclosed AI use did not work hard. P13, the government-school exception, accepted substantial assistance with quiz checking and described quiz-related disclosure as less sensitive than disclosure of personal feedback. For these teachers, a standard made AI marking checkable, and whether they delegated it depended on whether students expected marking to show the teacher's labour.

Encouragement and emotional support produced the most uniform codes: eleven teachers excluded AI. The standard here was knowledge of the student that no rubric contains. P6 ran a question-by-question AI analysis of student marks with red, amber and green coding, but said he would not let a marks-based forecast decide the grade of a student who had been in hospital, because it would fail a student whose illness the analysis could not see. P7 kept emotional conversations and parent meetings face-to-face. P1 tied this work to the teacher's role:

\begin{quote}
Teacher is not a robot. Teacher is a mentor. Students copy you\ldots{} If you use AI all the time, you become a robot. (P1, NGO school)
\end{quote}

\noindent P4 and P16, the two C codes on encouragement, show how knowledge of a student could become a usable standard. P4 supplied the personal information a message should convey before accepting draft wording, and P16 weighed AI suggestions against the student's situation.

\subsection{Checking Required the Textbook and Subject Expertise}
\label{sec:f-expertise}

The failures participants reported were failures of curriculum alignment. P17 generated a twelve-question paper and found that three or four questions matched the textbook. He kept using AI for test generation, coded D, and described cross-checking against the textbook as what turned AI from a risk into an asset. P18, a vice-principal, restricted teachers' use of ChatGPT after an AI-assisted answer contradicted the prescribed textbook. P9 warned that ChatGPT is ``not an O-level teacher'' and could produce explanations far beyond the depth the syllabus required. Outputs that were fluent and defensible on their own terms failed the teacher's test when they departed from the textbook on which the class would sit its examinations.

Comparing an output with that textbook required subject knowledge. P1 put the condition directly:

\begin{quote}
If you don't have the knowledge of the subject you're looking for answers for, it can be highly risky. (P1, NGO school)
\end{quote}

\noindent P4 learned this while preparing an unfamiliar IGCSE topic on differentiation. Two attempts to teach it with AI support failed, and the lesson worked only after P4 returned to the primary textbook. P4's rule afterwards was to read the primary text first and use AI once the topic was understood. P9 described the hazard this rule guards against: AI could give incorrect content a polished structure that a teacher without subject grounding would pass on.

Checking also extended to the class. P10 used AI to understand concepts and then re-expressed them for students who needed more explanation. P13 adapted AI-generated worksheets for pupils at different attainment levels and relied on Gemini's Urdu output in a multilingual rural classroom. Worksheets carry four unclear codes in Table~\ref{tab:joint}. Their standard combined the textbook with the teacher's knowledge of particular pupils, and teachers described adapting worksheets more readily than a fixed level of delegation.

\subsection{Institutions Resourced Checking Unequally}
\label{sec:f-institutions}

Teachers' accounts of the time available for checking differed by workplace. P3 taught classes of 50 to 70 pupils and described non-teaching assignments, including polio campaigns, elections and examination invigilation, that left a month of missed lessons. Describing teachers who return to such a backlog, P3 expected them to stop weighing the consequences of AI use:

\begin{quote}
We won't even consider whether the effect is negative or positive. (P3, government school)
\end{quote}

\noindent P3's own codes remained among the most restrictive in the sample, so the quotation describes an expectation about colleagues under backlog. P5's threshold, quoted above, came from a teacher with small classes and institutional tools.

Private-school teachers described organised support. P6 received a monthly staff award for incorporating AI into lesson planning and went on to lead regional training sessions, P7 mentored a colleague with three decades of experience in using ChatGPT, and P9's school provided built-in lesson-planning software. P7 also met colleagues who took reliance on ChatGPT as a sign that a teacher ``doesn't know anything.'' Government-school teachers described individual routes. P10 looked to departmental instructions before using AI more widely, P11 learned from peers and reported no departmental guidance, and P13 used Gemini frequently in a school where few colleagues did. P16 said he used AI ``like WhatsApp,'' informally and without a platform for learning professional uses.

Administrators placed checking with different people (Table~\ref{tab:governance}). P1 routed teachers' AI-assisted work through her own review and could recognise AI-generated lesson plans. P8 paired digital resources and staff training with an expectation that teachers cross-check outputs. P14 encouraged use through personal example without a formal policy, and P15 summarised his institute's position: ``there is no policy; everyone is using it for their own ease.'' P18 prohibited teacher use of ChatGPT, while teachers continued to use it informally and administrators used it for their own work. Where a coordinator reviewed work, part of the checking moved off the individual teacher; elsewhere it stayed with the teacher or became a prohibition. Each arrangement was local: the provincial department had issued no guidance on teachers' AI use (Section~\ref{sec:setting}). The theme ratings show the same divide. The coding record rated institutional support strong for four of five private-school teachers and none of the six government-school teachers, and rated the absence of formal policy strong for five of the six government-school teachers (Appendix Table~\ref{tab:themes}).

\begin{table}[t]
\caption{Local arrangements for AI use described by participants with oversight roles.}
\label{tab:governance}
\small
\begin{tabular}{@{}lp{3.6cm}p{2.5cm}@{}}
\toprule
ID & Reported arrangement & Who checks AI-assisted work\\
\midrule
P1 & Teachers generate ideas with AI and submit work for review & Coordinator\\
P8 & Digital books, apps, AI courses and staff training & Teachers, with institutional training\\
P14 & Principal models use of Gemini Pro; no formal policy & Individual teachers\\
P15 & Teachers use AI for papers, notes and content; no formal policy & Individual teachers\\
P18 & Teacher use of ChatGPT prohibited after a textbook contradiction & Administration\\
\bottomrule
\end{tabular}
\end{table}

\subsection{Disclosure Conveyed Whether the Teacher Reviewed the Work}
\label{sec:f-disclosure}

Teachers disagreed about disclosing AI-assisted feedback, and each position rested on a prediction about what students would infer. P4 took the most open position, ``I would always tell my students that I used AI,'' because students already recognised AI-generated text. P5 predicted the opposite reception:

\begin{quote}
Students will perceive it as all generative AI response and think it won't help them improve. (P5, private school)
\end{quote}

\noindent P5 nonetheless supported disclosure of rubric-based marking, where the teacher could explain the method. P6 accepted disclosure for assessment aligned with learning outcomes and resisted it for welcome notes and personal remarks meant to convey care. P1 favoured telling senior students and shielding younger ones, and expected parents who discovered AI-written messages to conclude that the school did not care about their child.

P2, the student participant, judged feedback by whether the teacher's own judgment was present in it. P2 accepted AI-generated wording when it carried the teacher's reading and assessment criteria, rejected feedback the teacher had pasted without reading, and described discovering undisclosed use as a reason to seek help from someone else while remaining motivated to learn. P5 expected students to discount AI involvement in itself; P2 discounted the absence of the teacher's review. Participants judged students' own use of AI on different grounds: the practice students needed in order to learn (P4, P7, P15).

\section{Discussion}
\label{sec:discussion}

Our participants decided which components of a task they could check, and against what.

\subsection{Standard-Anchored Delegation}
\label{sec:d-standard}

Assessment research describes judging work as holding a standard, comparing the work with it and acting on the difference~\cite{sadler1989formative}. Our participants applied the same logic to AI output. We call the resulting pattern \emph{standard-anchored delegation}: a teacher delegates a component of work when a standard outside the model defines what a correct output looks like and the teacher can state that standard and apply it to the result. The teacher may give the standard to the model before generation, as P5 did with a rubric (Section~\ref{sec:f-standard}), or apply it during review, as P17 did with the textbook (Section~\ref{sec:f-expertise}). Both routes depend on the teacher holding the standard and having the expertise and time to compare the output with it, the capability Tai et al.\ call evaluative judgement~\cite{tai2018evaluative}. P1, P4 and P9 described that expertise as the condition of safe use (Section~\ref{sec:f-expertise}).

The standards in our data took three forms. Codified standards such as a textbook, rubric or answer key are written down and easy to state. This is why preparation was the most delegated work, with nine teachers coded C or D on lesson planning and on quiz generation (Section~\ref{sec:f-production}), and why teachers who accepted AI marking tied it to a rubric (P5, P16; Section~\ref{sec:f-standard}). Knowledge of particular learners, one of the categories in Shulman's knowledge base for teaching~\cite{shulman1987knowledge}, is largely tacit. It became a usable standard when a teacher made it explicit, as P4 did by writing a student's circumstances into the request for encouragement, a conversion Nonaka calls externalisation~\cite{nonaka1994dynamic}. It could also be applied in review, as P16 did when weighing AI suggestions against a student's situation (Section~\ref{sec:f-standard}). Learning objectives occupy the place constructive alignment gives them: the outcomes from which activities and assessments are derived~\cite{biggs1996enhancing}. Objectives can themselves be checked against the curriculum, but they are the standard by which a teacher judges the rest of a lesson. No teacher in our sample accepted an AI draft of them, and P1 described keeping them while AI drafted the activities (Section~\ref{sec:f-production}).

The concept makes a testable claim. A teacher who holds a codified standard and can state it, yet withholds the component, is evidence that something other than verifiability governs the decision. Our data contain such cases. The government-school teachers who refused AI marking (P3, P12, P17) worked with quizzes that have answer keys, and they explained the refusal through the effort students expected to see (Section~\ref{sec:f-standard}). An available standard was therefore necessary for delegation in our sample and was not sufficient. Where a component's value to students lay partly in the teacher being seen to do it, teachers withheld it even when they could check it.

\subsection{Verifiability, Standards and Alignment}
\label{sec:d-constructs}

Lubars and Tan's factors of motivation, difficulty, risk and trust are judgments a person forms about a task~\cite{lubars2019ask}. A standard belongs to a component of the task, and its availability varies within a single activity: of the nine teachers who delegated lesson planning, eight kept AI to formatting or suggestions on the same plan's objectives (Section~\ref{sec:f-production}). Models of levels of automation already allow automation to vary across stages of information processing~\cite{parasuraman2000model}. The components that drew different codes, a lesson's activities and its objectives, sit at the same stage, and what separated them was whether the teacher held a standard for the output. The pattern fits Shneiderman's separation of automation from human control~\cite{shneiderman2020hcai}. Teachers accepted extensive AI production where they kept the authority to accept, revise or reject it, as P17 did when he kept generating question papers and checked each one against the textbook (Section~\ref{sec:f-expertise}).

The closest HCI construct is verifiability. Fok and Weld argue that explanations help only when they let people verify AI output, and that most tasks resist verification~\cite{fok2024verifiability}. Software testing names the same difficulty the oracle problem~\cite{barr2015oracle}. Accounts of appropriate reliance assume a correct answer against which people can calibrate reliance~\cite{lee2004trust,vasconcelos2023explanations}. In our setting, verifiability depended less on the output and its explanation than on whether the teacher held a standard and could bring it to the interaction. Teachers built a local stand-in for the correct answer from the textbook (P17, P18, P9) and from their knowledge of the class (P10, P13; Section~\ref{sec:f-expertise}).

Assessment and curriculum research supplies the other half of the account. Sadler's conditions for judging work~\cite{sadler1989formative} and the alignment of activities to outcomes~\cite{biggs1996enhancing,wiggins2005understanding} describe standards teachers apply to their own and their students' work. Our participants applied them to AI output, and they predicted where teachers drew boundaries within a task. In Table~\ref{tab:joint}, the categories with codified standards drew most of the C and D codes, while learning objectives and encouragement drew none and two respectively. Tankelevitch et al.\ describe the evaluation generative AI asks of its users~\cite{tankelevitch2024metacognitive}; teachers' subject knowledge and evaluative judgement determined whether they could carry it out, as P4's failed IGCSE lesson showed (Section~\ref{sec:f-expertise}).

Pakistan's textbook-bound, examination-driven schooling made codified standards unusually authoritative. P17, P18 and P9 judged outputs against the textbook a board examination would test (Section~\ref{sec:f-expertise}). We expect standard-anchored delegation to look different where curricula leave teachers more discretion: with fewer codified standards, more components would depend on knowledge of learners and on teacher-set objectives, which our participants were least willing to delegate.

\subsection{Care, Effort and the Limits of Verifiability}
\label{sec:d-care}

An alternative explanation for withheld encouragement is that teachers regarded care as authentic only when it came from them. Users of AI writing tools find them inauthentic in close relationships~\cite{fu2024text}, and P1's insistence that the teacher is a mentor fits this reading (Section~\ref{sec:f-standard}). The two accounts predict different behaviour once the teacher supplies the student's situation. An authenticity norm forbids AI wording regardless, while standard-anchored delegation permits it once the teacher holds the standard. P4 and P16 accepted wording after supplying or weighing the student's circumstances, which supports the latter account for them. The authenticity account better explains the eleven teachers who excluded AI from encouragement without considering such a procedure (Table~\ref{tab:joint}). The government-school refusals of AI marking show that expected effort operated even for components with codified standards. People judge work by the effort they believe it took~\cite{kruger2004effort}, and P3, P11 and P12 expected students to apply that judgement to marking (Section~\ref{sec:f-standard}). Verifiability marks what teachers could delegate; expectations of care and effort shaped what they chose to.

\subsection{Checking as Invisible Work}
\label{sec:d-labour}

If delegation depends on checking, then the time and expertise for checking determine what a teacher can safely delegate. Bainbridge observed that automation leaves operators the harder work of monitoring~\cite{bainbridge1983ironies}, and generative AI moves users from producing content to evaluating it~\cite{simkute2024ironies,lee2025critical}. Star and Strauss describe how work becomes invisible when the people who count it do not see it~\cite{star1999layers}. The EEF trial of ChatGPT for lesson preparation measured teachers' workload through a weekly diary and discussed guidance on checking the tool's outputs~\cite{roy2024chatgpt}. Our participants treated that checking as the condition of use. Histories of educational automation record the same pattern of unrecognised teacher repair work~\cite{bergvikenrensfeldt2023automating,selwyn2024limits}. In our sample the invisible work fell unevenly (Section~\ref{sec:f-institutions}). P1's coordinator review and Shiksha Copilot's curators~\cite{dennison2026shiksha} each take part of the check off the classroom teacher, and P5, P6 and P9 worked with institutional planning tools. Most government-school teachers, with the largest classes and the most non-teaching duties~\cite{nasreen2019time}, checked alone, and P3 expected colleagues returning to a backlog of missed lessons to stop weighing the consequences of AI use.

\subsection{Disclosure as a Report of Review}
\label{sec:d-disclosure}

Studies of AI-mediated communication find that disclosed AI involvement lowers trust in a sender and reads as low effort~\cite{jakesch2019aimc,liu2022console}. Our participants agreed that recipients would read disclosure as a statement about effort, and they specified which effort mattered (Section~\ref{sec:f-disclosure}). P5 and P11 anticipated that students would discount AI involvement, and P2 described discounting feedback the teacher had not read. Review is effort that recipients cannot observe directly. Buell and Norton show that people value a service more when they can see the work behind it~\cite{buell2011labor}. A bare AI label hides the teacher's review, while a statement of what the teacher checked and changed shows it to the recipient. P5's support for disclosing rubric-based marking, where the teacher could explain the method, fits this reading. The statement reports the review; a logged edit documents part of it, and the soundness of the assessment remains the teacher's responsibility. This extends Draxler et al.'s distinction between ownership and declared authorship~\cite{draxler2024ghostwriter}: for teachers, the relevant declaration concerns review.

\subsection{Limitations and Future Work}
Student perspectives only came from one participant, and we did not interview parents; experiments with students and parents should test whether disclosure that reports a teacher's review changes trust. Participants reported their practices and reasoned live about scenarios. Deployment studies that log generation alongside teachers' edits would measure how much checking occurs and whether standard-anchored delegation predicts what teachers hand over.

\section{Implications for Design and Policy}
\label{sec:implications}

Standard-anchored delegation implies that designers should evaluate teacher-facing AI by how well it supports checking, and that institutions adopting it take on the cost of that checking. Each recommendation below follows from a kind of standard in Table~\ref{tab:joint}.

\subsection{Teacher-Facing AI Systems}

\paragraph{R1: Ask for the codified standard before generating.} Tools for lesson planning and question papers should require the board, grade, textbook and chapter, and should let the teacher fix the lesson's objectives or the rubric before generation. The same request should carry the language of instruction and the attainment level of the class, which P13 and P10 otherwise adjusted by hand. Teachers who delegated successfully already supplied these standards; the interface should make supplying them the default.

\paragraph{R2: Show each output beside its standard.} Each generated question or explanation should appear next to the textbook passage it draws on, so that the comparison P17 made by hand becomes a visual check. This gives Amershi et al.'s guidance on efficient correction~\cite{amershi2019guidelines} a concrete object: the teacher corrects toward a named curricular document, which supplies the correct answer the model lacks.

\paragraph{R3: Turn review into a reportable statement.} For feedback on student work, the system should log the criteria the teacher applied and the comments the teacher changed, and draft a statement for the teacher to confirm, such as: \emph{``I marked this against the Chapter 6 rubric and rewrote the comments on questions 3 and 5.''} The teacher confirms or corrects the statement and decides whether to share it. The statement reports the review P2 sought, and the logged edits document the changes it describes. With teachers' consent, the same edits record which AI comments teachers rewrote, the kind of signal Woodrow et al.\ used to train feedback models with teachers in the loop~\cite{woodrow2025feedback}.

\subsection{Schools and Education Departments}

\paragraph{R4: Match oversight to the kind of standard.} School guidance should let teachers self-check components with codified standards, such as question papers drawn from the prescribed textbook. Components that depend on knowledge of learners, such as encouragement and parent messages, should stay under the authority of the teacher who knows the student. A teacher may use AI wording for these messages, as P4 and P16 did, once the teacher has supplied the student's circumstances and approved the result. A term's learning objectives suit review by a coordinator or subject head, as in P1's school. A review requirement scoped this way addresses the risk that led P18's college to prohibit teacher use, while keeping teachers' access to planning support.

\paragraph{R5: Budget checking time and write rules for teachers' own use.} Departments introducing AI tools should count the time teachers spend comparing outputs with the textbook and revising feedback as part of their workload, and should support subject-based peer groups for sharing curriculum-checked prompts, building on the WhatsApp channels Hafeez and Zehra used for teacher support~\cite{hafeez2025assistant}. The national AI policy and the examination boards' rules address capacity building and students' use (Section~\ref{sec:setting}), and the provincial department had issued no guidance for teachers at the time of our study. School guidance should add rules for teachers' feedback and parent communication that state when teachers must disclose AI assistance and require the disclosure to describe the teacher's review.

\section{Conclusion}

Teachers in Punjab delegated to generative AI the components of their work they could check, and kept the standards against which they checked them. Teachers delegated lesson activities widely and tied AI marking to rubrics, both checkable against codified standards; they withheld learning objectives and encouragement, which depended on goals they set and on their knowledge of students. An available standard was necessary for delegation without being sufficient, since government-school teachers withheld marking they could check when students expected to see their effort. Checking depended on subject expertise and on time that institutions allocated unequally, and participants read disclosure as a report of that checking. Teacher-facing AI and school policy can act on each of these findings by asking for replicable standards, displaying them, reporting review and budgeting the time it takes.

\begin{acks}
\textbf{Use of generative AI.} In accordance with the ACM Policy on Authorship, we disclose the following uses of AI tools. Interview recordings were transcribed with automated multilingual speech-recognition models. During the writing of this paper, we used Claude (Anthropic; Claude Opus 5.5) to  compile the appendix summary tables from our coding records, and check citation details and the consistency of reported counts. The authors reviewed and edited all AI-assisted content, and take full responsibility for the content of this paper.
\end{acks}

\bibliographystyle{ACM-Reference-Format}
\bibliography{refs}

\appendix
\section{Teacher Task Codes}
\label{app:codes}

\begin{table*}[t]
\caption{Themes rated strong in the cross-case coding record, by participant group. Group cells give the number rated strong over the group size, with participant IDs; single-participant cells give the rating (S = strong, M = moderate, W = weak). Admin.\ = administrators (P8, P14, P15, P18). Section gives where the paper reports the theme; -- = not analysed in this paper.}
\label{tab:themes}
\footnotesize
\setlength{\tabcolsep}{3pt}
\begin{tabular}{@{}p{4.1cm}lp{2.7cm}p{2.3cm}ccp{2.3cm}cc@{}}
\toprule
Theme & Section & Gov.\ teachers ($n=6$) & Priv.\ teachers ($n=5$) & P1 & P16 & Admin.\ ($n=4$) & P2 & All ($n=18$)\\
\midrule
Delegating some components of a task and keeping others & \ref{sec:f-production}--\ref{sec:f-standard} & 4/6 (P3, P12, P13, P17) & 5/5 & S & S & 4/4 & M & 15\\
Teacher review as a condition of use & \ref{sec:f-standard}--\ref{sec:f-expertise} & 6/6 & 5/5 & S & S & 4/4 & S & 18\\
Relational work kept by the teacher & \ref{sec:f-standard} & 4/6 (P3, P12, P13, P17) & 3/5 (P5, P6, P7) & S & M & 2/4 (P8, P15) & M & 10\\
Textbook and syllabus alignment & \ref{sec:f-expertise} & 3/6 (P3, P13, P17) & 3/5 (P5, P6, P9) & W & M & 2/4 (P14, P18) & W & 8\\
Urdu and localisation & \ref{sec:f-expertise} & 1/6 (P13) & 1/5 (P7) & W & W & 0/4 & W & 2\\
Workload pressure & \ref{sec:f-institutions} & 4/6 (P3, P12, P13, P17) & 1/5 (P4) & S & M & 1/4 (P14) & W & 7\\
Institutional support for AI use & \ref{sec:f-institutions} & 0/6 & 4/5 (P5, P6, P7, P9) & M & W & 2/4 (P8, P14) & W & 6\\
Absence of formal policy & \ref{sec:f-institutions} & 5/6 (P3, P10, P11, P12, P17) & 1/5 (P4) & M & S & 3/4 (P14, P15, P18) & W & 10\\
Training and tool literacy & \ref{sec:f-institutions} & 4/6 (P3, P10, P11, P17) & 5/5 & M & S & 4/4 & W & 14\\
Government or rural constraints & \ref{sec:f-institutions} & 6/6 & 0/5 & M & M & 0/4 & W & 6\\
Disclosure and trust & \ref{sec:f-disclosure} & 4/6 (P11, P12, P13, P17) & 4/5 (P4, P5, P6, P7) & S & W & 1/4 (P18) & S & 11\\
Parents' trust & \ref{sec:f-disclosure} & 1/6 (P13) & 1/5 (P4) & S & W & 2/4 (P8, P18) & W & 5\\
Students' own use and misuse & \ref{sec:f-disclosure} & 1/6 (P17) & 3/5 (P4, P7, P9) & W & W & 1/4 (P15) & M & 5\\
Experience or generational differences & -- & 4/6 (P3, P10, P11, P13) & 2/5 (P7, P9) & W & S & 3/4 (P14, P15, P18) & W & 10\\
\bottomrule
\end{tabular}
\end{table*}

Table~\ref{tab:codes} reports each teacher's code for the twelve task categories. Codes follow Section~\ref{sec:method}: A = no AI; B = ideas or checking only; C = draft edited or reviewed; D = substantial or near-final assistance; ? = unclear. P13's parent-communication code is mixed because P13 described different positions for different messages.

\begin{table}[htbp]
\caption{Teacher task codes by participant. LP = lesson planning; QG = quiz and test generation; GR = quiz grading; WS = worksheets and differentiation; RC = report-card comments; FB = student feedback; PC = parent communication; LO = learning objectives; EN = encouragement and emotional support; DA = data analysis and monitoring; PR = presentations and content; PT = policy and training.}
\label{tab:codes}
\footnotesize
\setlength{\tabcolsep}{2.6pt}
\begin{tabular}{@{}llcccccccccccc@{}}
\toprule
ID & Sector & LP & QG & GR & WS & RC & FB & PC & LO & EN & DA & PR & PT\\
\midrule
P1 & NGO & C & C & B & C & C & C & C & A & A & C & C & C\\
P3 & Gov. & B & C & A & B & ? & ? & ? & B & A & B & B & C\\
P4 & Priv. & D & B & C & C & C & C & C & B & C & C & C & B\\
P5 & Priv. & D & D & C & C & C & C & C & B & A & B & C & C\\
P6 & Priv. & D & C & C & D & B & C & B & B & A & D & D & C\\
P7 & Priv. & C & C & C & C & B & C & A & B & A & B & C & B\\
P9 & Priv. & D & C & B & B & ? & B & ? & B & A & B & D & C\\
P10 & Gov. & B & B & ? & ? & ? & ? & ? & B & A & B & B & B\\
P11 & Gov. & B & C & B & ? & ? & ? & ? & ? & A & ? & B & B\\
P12 & Gov. & B & B & A & C & C & A & B & B & A & B & B & B\\
P13 & Gov. & D & B & D & C & A & ? & A/D & B & A & B & B & B\\
P16 & Voc. & C & D & C & ? & ? & C & ? & ? & C & ? & C & B\\
P17 & Gov. & C & D & A & ? & ? & A & A & B & A & B & B & B\\
\bottomrule
\end{tabular}
\end{table}

Table~\ref{tab:sector} summarises the codes by sector for the 13 teachers. Table~\ref{tab:themes} reports, for all 18 participants, the themes the coding record rated as strong. A count such as 4/6 means that four of the six participants in the group received a strong rating; the remaining participants received moderate or weak ratings, which do not indicate absence.

\begin{table}[htbp]
\caption{Task codes by sector (teachers only). Government ($n=6$) and private ($n=5$) columns give the number of teachers coded C or D, coded A, and unclear (?). P1 (NGO) and P16 (vocational) show their individual codes. One government parent-communication code is mixed A/D (P13) and is counted in neither column.}
\label{tab:sector}
\footnotesize
\setlength{\tabcolsep}{3pt}
\begin{tabular}{@{}lrrrrrrcc@{}}
\toprule
 & \multicolumn{3}{c}{Gov.\ ($n=6$)} & \multicolumn{3}{c}{Priv.\ ($n=5$)} & & \\
\cmidrule(lr){2-4}\cmidrule(lr){5-7}
Task category & C/D & A & ? & C/D & A & ? & P1 & P16\\
\midrule
Lesson planning & 2 & 0 & 0 & 5 & 0 & 0 & C & C\\
Quiz and test generation & 3 & 0 & 0 & 4 & 0 & 0 & C & D\\
Presentations and content & 0 & 0 & 0 & 5 & 0 & 0 & C & C\\
Worksheets and differentiation & 2 & 0 & 3 & 4 & 0 & 0 & C & ?\\
Quiz grading & 1 & 3 & 1 & 4 & 0 & 0 & B & C\\
Student feedback & 0 & 2 & 4 & 4 & 0 & 0 & C & C\\
Report-card comments & 1 & 1 & 4 & 2 & 0 & 1 & C & ?\\
Parent communication & 0 & 1 & 3 & 2 & 1 & 1 & C & ?\\
Encouragement and support & 0 & 6 & 0 & 1 & 4 & 0 & A & C\\
Learning objectives & 0 & 0 & 1 & 0 & 0 & 0 & A & ?\\
Data analysis and monitoring & 0 & 0 & 1 & 2 & 0 & 0 & C & ?\\
Policy and training & 1 & 0 & 0 & 3 & 0 & 0 & C & B\\
\bottomrule
\end{tabular}
\end{table}

\section{Interview Materials}
\label{app:protocol}

The eleven teacher task prompts were: generating a short quiz; making a vocabulary list or summary; formatting lesson objectives; suggesting worksheet ideas for mixed-ability students; summarising class performance trends; drafting report-card comments; drafting feedback on a weak essay; drafting feedback for a struggling student who is losing confidence; writing a parent message about academic performance; writing a parent message about behaviour; and suggesting extra support for a learner. For each prompt we asked about comfort with AI use, the intended form of assistance, reasons for declining it, possible problems, the people affected and the checking or editing required. The administrator protocol covered teacher workload, the institution's formal or informal stance on AI, acceptable and risky uses, sensitive communication, disclosure and differences between settings. The student protocol covered meaningful feedback, perceived teacher effort, awareness of AI, reactions to the two disclosure vignettes and other sources of academic support.

\end{document}